\documentclass[conference]{IEEEtran}
\usepackage{cite}
\usepackage{amsmath,amssymb,amsfonts}
\usepackage{algorithmic}
\usepackage{graphicx}
\usepackage{textcomp}
\usepackage{xcolor}
\def\BibTeX{{\rm B\kern-.05em{\sc i\kern-.025em b}\kern-.08em
    T\kern-.1667em\lower.7ex\hbox{E}\kern-.125emX}}

\usepackage{lipsum}
\usepackage{url}
\usepackage{booktabs}
\usepackage{multirow}
\usepackage{xcolor}
\usepackage{makecell}
\usepackage{enumitem}

\usepackage{hyperref}
\usepackage{cleveref}

\usepackage{subcaption}

\usepackage[most]{tcolorbox}

\newtcolorbox{keytakeawaybox}{
  colback=green!10,    
  colframe=gray!50,
  boxrule=1pt,          
  arc=1pt,              
  left=3pt,             
  right=3pt,            
  top=3pt,              
  bottom=3pt,            
  before skip = 1ex, 
}

\newcounter{keytakeawayid}
\newcommand{\defkeytakeaway}[1]{\refstepcounter{keytakeawayid}\label{#1}\textbf{Key Takeaway \arabic{keytakeawayid}:}}

\newcommand{\changes}[1]{#1}

\newif\ifarXiv
\arXivtrue

\newcommand{\toolname}[0]{STEAM}

\newcommand{\encircle}[1]{%
	\begin{tikzpicture}[baseline=(char.base)]
		\node[draw=gray,circle,inner sep=0.5pt, fill=gray, text=white] (char){\small #1};
	\end{tikzpicture}%
}

\newcommand{\encirclered}[1]{%
	\begin{tikzpicture}[baseline=(char.base)]
		\node[draw=black,circle,inner sep=0.5pt, fill=red, text=white] (char){\small #1};
	\end{tikzpicture}%
}

\newcommand{\encircleblue}[1]{%
	\begin{tikzpicture}[baseline=(char.base)]
		\node[draw=black,circle,inner sep=0.5pt, fill=blue, text=white] (char){\small #1};
	\end{tikzpicture}%
}

\begin{document}

\title{OpenDC-STEAM: Realistic Modeling and Systematic Exploration of Composable Techniques for Sustainable Datacenters\\
\large Technical Report
}

\author{
\IEEEauthorblockN{Dante Niewenhuis}
\IEEEauthorblockA{\textit{Vrije Universiteit Amsterdam}\\
\textit{d.niewenhuis@vu.nl}}
\and
\IEEEauthorblockN{Sacheendra Talluri}
\IEEEauthorblockA{\textit{Vrije Universiteit Amsterdam}\\
\textit{s.talluri@vu.nl}}
\and
\IEEEauthorblockN{Alexandru Iosup}
\IEEEauthorblockA{\textit{Vrije Universiteit Amsterdam}\\
\textit{a.iosup@vu.nl}}
\and
\IEEEauthorblockN{Tiziano De Matteis}
\IEEEauthorblockA{\textit{Vrije Universiteit Amsterdam}\\
\textit{t.de.matteis@vu.nl}}
}

\maketitle

\begin{abstract}
The need to reduce datacenter carbon-footprint is urgent. 
While many sustainability techniques have been proposed, they are often evaluated in isolation, using limited setups or analytical models that overlook real-world dynamics and interactions between methods. This makes it challenging for researchers and operators to understand the effectiveness and trade-offs of combining such techniques.
%
We design OpenDC-\toolname{}, an open-source customizable datacenter simulator, to investigate the individual and combined impact of sustainability techniques on datacenter operational and embodied carbon emissions, and their trade-off with performance. 
%
Using \toolname{}, we 
systematically explore three representative techniques--horizontal scaling, leveraging batteries, and temporal shifting--with diverse representative workloads, datacenter configurations, and carbon-intensity traces. 
Our analysis highlights that datacenter dynamics can influence their effectiveness and that combining strategies can significantly lower emissions, but introduces complex cost-emissions-performance trade-offs that \toolname{} can help navigate.
\toolname{} supports the integration of new models and techniques, making it a foundation framework for holistic, quantitative, and reproducible research in sustainable computing. 
Following open-science principles, \toolname{} is available as FOSS: \url{https://github.com/atlarge-research/OpenDC-STEAM}.
\ifarXiv
This is an extended version of a paper published at CCGRID 2026.
\fi
\end{abstract}


\section{Introduction}\label{sec:introduction}

The Information and Communication Technology (ICT) sector accounts for an estimated 1.5\%-4\% of global carbon emissions~\cite{belkhir2018assessing, freitag_ict_emissions},  
with 
datacenters contributing for a large and growing share  \cite{acm2021computing, wilson2023era}.
Policy mandates, user expectations, and financial considerations are pushing the computing community to control and reduce datacenter carbon footprint~\cite{cacm_our_house_is_on_fire, cacm_chien_computing_grand_challenge}.
In response, many techniques have been proposed to reduce the carbon emissions of a datacenter, including load shifting~\cite{acun2023carbon, sukprasert2024limitations}, hardware scaling~\cite{lin2012dynamic,  uddin2010server}, or batteries~\cite{acun2023carbon, sheme2018battery, li2017balancing}.
%
However, these strategies are often evaluated using analytical models or limited setups that do not account for interactions between sustainability techniques and with real-world datacenter dynamics, such as task failures.
Their effectiveness also depends on multiple factors, such as the scale of the datacenter, workload characteristics, the energy source \textit{carbon intensity}, and the \textit{embodied carbon} emitted during hardware manufacturing~\cite{sukprasert2024limitations, niewenhuis2024footprinter}. In such a broad design space, non-trivial trade-offs in costs, carbon emissions, and performance frequently arise.
To navigate this complexity and make informed decisions on how to design datacenters and configure their operations, \textit{datacenter designers, operators, and sustainability researchers need quantitative insights}. 
In this work, we introduce \textbf{OpenDC-\toolname{}} (\toolname{} in the following), an open-source simulation-based tool for quantifying the effects of sustainability techniques on datacenter performance and carbon footprint. 

\begin{figure}[t]
    \centering
    \includegraphics[width=\linewidth]{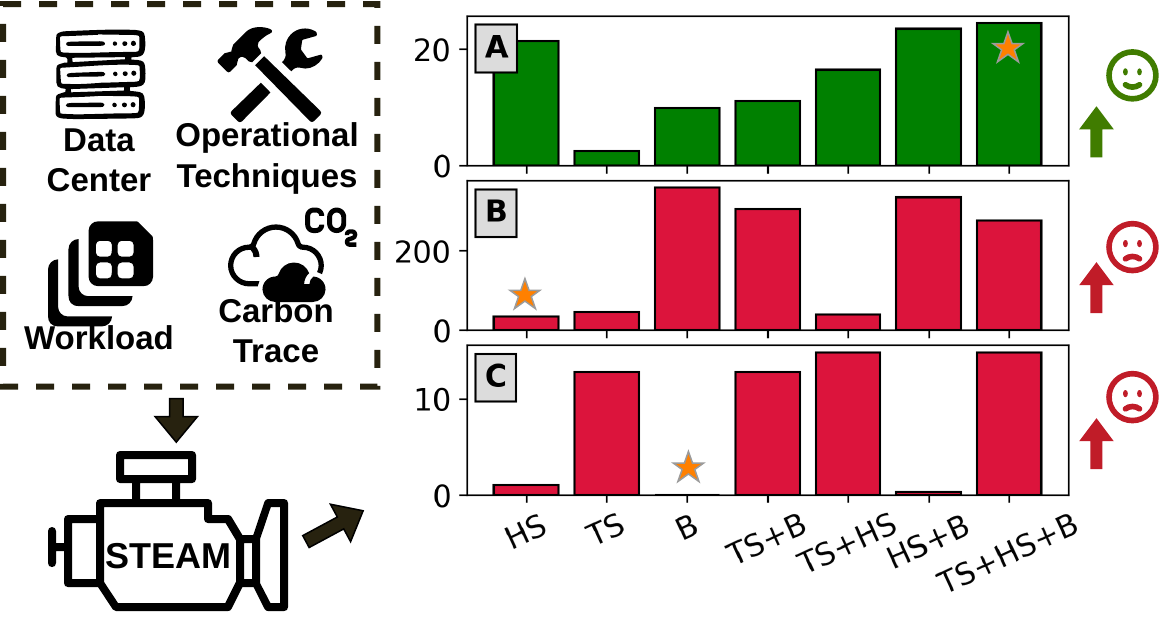}
    \caption{\toolname{} quantifies the impact and trade-offs of sustainability techniques. In the figure, (A) results for the Surf workload are shown as total carbon reduction~[\%], (B) peak power draw~[kW], and (C) average task delay~[h]. Stars indicate the best performing technique for each metric. HS: Horizontal Scaling, TS: Temporal Shifting, B: Batteries. }
    \label{fig:first_page}
    
\end{figure}

    

\toolname{} is a new capability-driven feature in the OpenDC space~\cite{mastenbroek2021opendc}.
\toolname{}'s composable architecture allows users to analyze trade-offs arising from combining techniques, evaluate new techniques, and tune existing ones. Its flexibility and performance enable datacenter designers, operators, and researchers to rapidly explore a wide range of "what-if" scenarios and gain actionable insights into sustainability decisions. 

We use \toolname{} to quantitatively evaluate three representative sustainability techniques--horizontal scaling, temporal shifting, and batteries--on datacenter performance (e.g., service quality) and carbon footprint, considering diverse and realistic workloads, carbon regions, datacenters, and their dynamics. We show that each technique can reduce carbon emissions with minimal performance impact, but the actual extent of this reduction depends on the workload, available hardware, location, and operational phenomena such as task failures. We highlight how disregarding datacenter dynamics has led to optimistic carbon emissions reductions in prior work.
\Cref{fig:first_page} illustrates how \toolname{} can be used to understand the impact of sustainability techniques. Combining techniques can further improve their effectiveness (see \Cref{fig:first_page}A), but also introduces trade-offs, such as power-draw spikes (\Cref{fig:first_page}B) and service-quality degradation (\Cref{fig:first_page}C), which \toolname{} can quantify.

\changes{
With \toolname{}, we contribute to the efforts towards better understanding modern computing ecosystems\cite{Iosup2018MassivizingCS}. \toolname{} provides a framework for the systematic evaluation of sustainability techniques, helping datacenter providers and researchers to make evidence-based decisions on datacenter design and operation.
}

This work makes the following contributions:
\begin{enumerate}
    \item We motivate and design \toolname{}\footnote{\url{https://github.com/atlarge-research/OpenDC-STEAM}}~($\S$\ref{sec:motivation} and $\S$\ref{sec:steam}). \toolname{} is an open-source, flexible framework for quantifying the effect of sustainability techniques on the datacenter carbon footprint and performance. 
    
    \item Using \toolname{} through a relevant experimental setup~($\S$\ref{sec:setup}), we conduct an extensive analysis of three representative sustainability techniques, showing that:
    \begin{enumerate}
        \item Horizontal scaling can reduce carbon footprint (up to 35\%), but ignoring operational phenomena, such as failure, leads to optimistic results~($\S$\ref{subsec:horizontal_scaling}).
        \item Batteries can significantly improve the sustainability of datacenters (up to 28\%) when used in the right context, but can also reduce it when misused ~($\S$\ref{subsec:BAT}).
        \item Temporal shifting is less effective (7\%) than shown by previous work when operational effects, such as stacking tasks, are taken into account~($\S$\ref{subsec:temporal_shifting}).
    \end{enumerate}   
    \item We leveraged \toolname{}'s composable design to conduct a first-of-its-kind analysis of using multiple sustainability techniques in conjunction~($\S$\ref{sec:exp_combined}). Combining techniques can increase effectiveness but add new, complex trade-offs that \toolname{} helps quantify and explain.
\end{enumerate}


\section{Background}\label{sec:model}

This section introduces the concepts and techniques that enable realistic modeling of sustainability techniques in \toolname{}. 

\subsection{Workload}\label{subsec:workload}
A workload contains all jobs executed by a datacenter in a specific period. We use a bag-of-tasks workload model where users submit jobs to a datacenter, each containing one or more tasks. 
Workloads are defined using traces containing the arrival times of all tasks, resource requirements, durations, deadlines, and sampled resource utilization over time. Workloads are often created by collecting operation logs of existing datacenters. 

\subsection{Carbon Emissions}\label{subsec:carbon_emission}
Carbon emissions are all the CO2 and CO2-equivalent of other greenhouse gases emitted by a datacenter.
We distinguish between \emph{operational} and \emph{embodied} carbon emissions.

\textit{Operational Carbon} 
refers to all emissions generated from producing energy used to power servers, cooling systems, and other infrastructure in a datacenter. The type of energy a datacenter uses significantly impacts its operational carbon footprint, with coal-based electricity emitting up to 20x more carbon than renewable sources~\cite{Gupta2022act}.  Since green energy availability fluctuates~\cite{sukprasert2024limitations}, accurate estimation requires information about the carbon intensity over time, often provided as traces from services such as ElectricityMaps\footnote{\url{https://www.electricitymaps.com/}} or Watttime\footnote{\url{https://watttime.org/}}.

\textit{Embodied Carbon}
considers emissions from manufacturing datacenter components such as CPUs and GPUs. As the energy efficiency of datacenters improves, embodied carbon has become a major contributor to the overall carbon footprint~\cite{lyu2023myths}. 
A common method to calculate the embodied carbon of a component during a workload is to multiply the total embodied carbon cost of the component by the portion of the component's lifetime the workload used~\cite{Gupta2022act}.

\subsection{Sustainability Techniques}\label{subsec:techniques}
Many techniques have been proposed to improve datacenter sustainability. 
In this work, we focus on techniques aimed at reducing the overall carbon emissions associated with the datacenter deployment and use. We consider three representative techniques: horizontal scaling, the use of batteries, temporal shifting, and their combination.

\textit{Horizontal scaling} reduces the number of machines in a datacenter to the minimum needed to meet computational requirements, decreasing energy usage and carbon emissions. Datacenters often run at very low utilization~\cite{shehabi2018data} as fears of service disruption have led to conservative practices and significant over-provisioning of resources~\cite{Glanz2012}. Although this might protect datacenters from sudden resource requirement spikes or failures, the idle machines still use energy and contribute to operational and embodied carbon emissions.

\textit{Batteries} can be used to reduce carbon emissions by storing green energy when available, and providing it to the datacenter in periods of high-carbon energy. Their advantage is that they reduce emissions without impacting task execution time. However, they also introduce downsides: manufacturing batteries has a high embodied carbon cost, and charging them can cause spikes in the datacenter's power draw (see \Cref{subsub:BAT:capacity_charging}).

\textit{Temporal shifting} delays or pauses a task during high-carbon periods, resuming it when greener energy is available. By moving tasks and, in turn, energy usage, to periods with lower carbon intensity, the overall operational carbon emitted is reduced. However, this requires tasks to be interruptable and delay-tolerant~\cite{sukprasert2024limitations}, and shifting many tasks to the same period might introduce workload spikes, requiring larger-scale datacenters to handle (see \Cref{sec:exp_combined}).
\section{Rationale for a Simulation-based Approach}\label{sec:motivation}
Datacenters are large and complex ecosystems composed of many interacting components. Evaluating the impact of sustainability techniques requires datacenter designers, operators, and researchers to account for all these components and their interactions.
Prior work often used single-purpose analytical models to estimate carbon savings. While useful for theoretical insights, such models struggle to capture datacenter dynamics and cross-technique interactions. To address this gap, we advocate a simulation-based approach. Simulation can enable a more comprehensive analysis by capturing datacenter operational phenomena and considering how different strategies interact with each other and the rest of the system.
In this section, we illustrate this using a motivating example concerning the analysis of the temporal shifting technique.  

Previous works~\cite{sukprasert2024limitations, bostandoost2024data} evaluated temporal shifting using analytical models similar to the following: for each task, emissions are calculated at its original execution time, and when delayed by a carbon-aware scheduling policy. The difference between those two represents the task's emission savings. The overall impact of temporal shifting is calculated by averaging the savings across all tasks.
This approach has two potential downsides. First, it assumes each task can be moved to a low-carbon period independently, ignoring the datacenter's computational capacity constraints. Although similar considerations can be integrated into the scheduling strategy, their impact is difficult to capture analytically. Moreover, the approach disregards the energy used by idle machines when a task is delayed. Second, the method overlooks operational phenomena such as failures. As a result, the analytical model might overestimate the carbon savings of temporal shifting.

\begin{figure}[t]
    \centering
    \includegraphics[width=\linewidth]{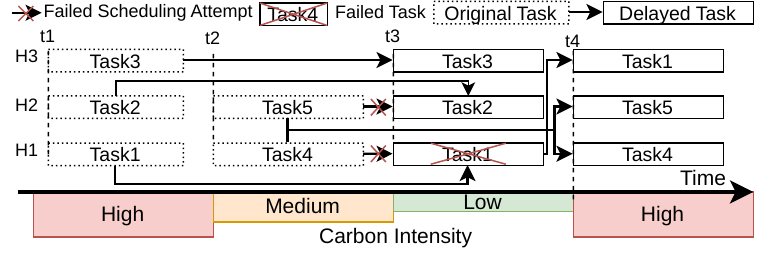}
    \caption{Example of a scenario challenging for analytical models to evaluate correctly.}
    \label{fig:analytic_shortcomings}
    \vspace*{-0.35cm}
\end{figure}

We illustrate the downsides with a simple example depicted in \Cref{fig:analytic_shortcomings}. The example concerns the operations of a datacenter containing three hosts (H1-3) that uses temporal shifting to reduce carbon emissions. The top of the figure shows the tasks submitted and scheduled on the different hosts. Dotted boxes indicate delayed tasks, with arrows pointing to the new scheduled time. A task marked with a red cross indicates it has been interrupted due to machine failure. The bottom of the figure depicts the carbon intensity over time.
We zoom into four timestamps, marked by $t1$-$t4$. At timestamp $t1$, three tasks (1-3) are submitted. Due to the high carbon intensity, the scheduling policy delays all tasks to $t3$. At timestamp $t2$, two new tasks (4,5) are submitted. Again, the scheduling policy delays the tasks to $t3$. At timestamp $t3$, no new tasks are submitted, but all five tasks are scheduled to start. Because only three tasks can run in parallel, two tasks (4,5) are delayed further to $t4$. Furthermore, during execution, host H1 fails, interrupting Task 1, causing it to be rescheduled at $t4$.

In this example, all five tasks are delayed by the scheduling policy. Using the analytical model discussed previously, it would be concluded that three tasks~(1-3) are successfully shifted from high to low carbon periods, and two tasks~(4,5) from medium to low carbon periods. In reality, however, only two tasks~(2, 3) are shifted successfully, with the other tasks~(1, 4, 5) being executed in a high-carbon period.
Although analytical models can be extended to model also failures, they still struggle to capture the actual behavior of a specific workload on a given datacenter, especially when multiple operational phenomena and sustainability techniques interact.
The composable simulator approach we propose addresses these challenges. In \toolname{}, each sustainability technique can be modeled by the user independently. They will then interact automatically with other techniques and datacenter dynamics during simulation, providing an accurate and quantitative view of their combined impact.

\section{\toolname{} Design}\label{sec:steam}

\toolname{} is a discrete event simulator, based on the OpenDC framework~\cite{mastenbroek2021opendc}, designed to evaluate the efficacy of sustainability techniques. It significantly extends OpenDC by adding support for operational and embodied carbon modeling, sustainability techniques, and GPUs. Notably, \toolname{} proposes a major redesign of OpenDC's core architecture to enable a composable and extensible simulator in which techniques can be independently developed and freely combined based on users' needs (see \Cref{sec:steam:abstractions}). This flexibility makes \toolname{} a powerful and versatile tool for holistic and quantitative research in sustainable datacenter design and operations.

\begin{figure}[t]
    \centering
    \includegraphics[width=\linewidth]{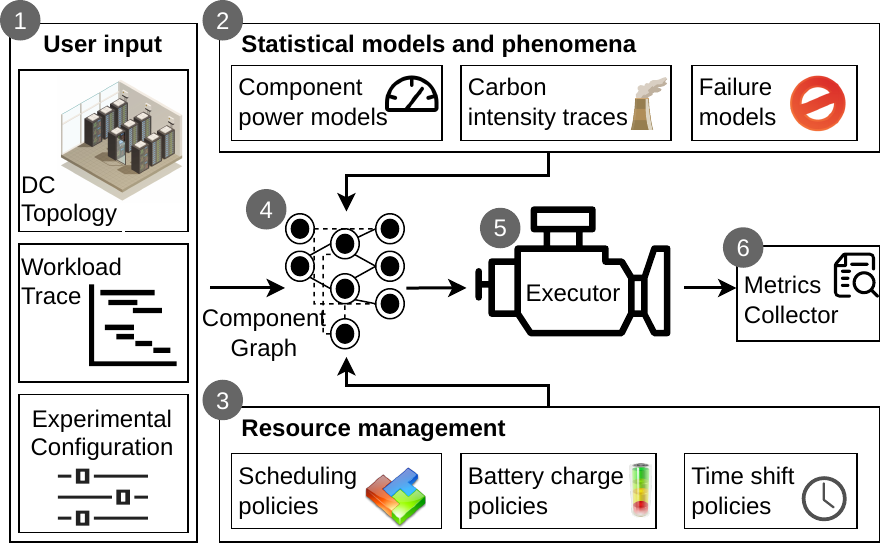}
    \caption{STEAM system architecture.}
    \label{fig:steam-overview}
    \vspace*{-0.35cm}
\end{figure}

\toolname{}'s architecture is depicted in \Cref{fig:steam-overview}. \toolname{} takes a datacenter topology, workload trace, and other experiment configurations as input~\encircle{1}. \toolname{} comes with ready-to-use statistical models~\encircle{2} the user can attach to components in the datacenter topology for use in simulation, such as power models for datacenter components (e.g., CPUs and GPUs), carbon intensity traces, and failure models. It also provides resource management primitives~\encircle{3} the user can configure to manage scheduling, battery management, and temporal shifting in the datacenter. Based on the user configuration, the datacenter components, respective statistical models, and resource managers are assembled into a component graph~\encircle{4}. The component graph is then executed by the event-based executor~\encircle{5}. The metrics collector~\encircle{6} samples the simulation at periodic intervals and collects data for analysis. The user can select which metrics to export and at which granularity.

\subsection{Overview of \toolname{} Components}
The user specifies the workload trace, the datacenter topology, carbon intensity traces, failure models, and configuration parameters such as a random seed for reproducibility. STEAM supports many common distributed systems workload formats.


The \emph{topology}~(part of~\encircle{1}) describes the available hardware in a datacenter. The user defines host types by providing CPUs, GPUs, and memory specifications. A user can provide different host types and their number. 
The user defines the power sources the hosts are connected to. Carbon traces can be connected to power sources to measure operational carbon. Users can add batteries to the datacenters. Batteries can be charged and discharged based on battery policies. 
Particularly relevant to this work are the CPU and GPU statistical power models (part of~\encircle{2}), which convert component utilization (in~\%) into power demand (in Watts). The user can calibrate and use built-in power models, including linear, square-root, square, and cubic. The user can also implement custom statistical power models.

Resource managers~\encircle{3} and operational techniques are needed to execute a workload on a datacenter. \toolname{} provides resource managers and operational techniques that support different scheduling algorithms and policies. The user can associate a resource manager with each resource, such as power, battery, compute, and memory. The user can specify scheduling policies in the experiment configuration (part of~\encircle{1}) using custom filters and weighers.


In \toolname{}, we implement various sustainability-focused resource management and operational techniques. These include threshold-based battery charge and discharge policies inspired by similar policies for batteries~\cite{sheme2018battery,li2017balancing}, as well as load shifting strategies~\cite{bostandoost2024data}. We implement task-delaying policies to shift tasks into lower-carbon-intensity periods, and a task-stopping mechanism that stops execution during high-carbon periods, resuming once greener energy becomes available. Users can easily configure and customize these policies with minimal configuration changes.

\subsection{Component Graph, Events, and Composability}\label{sec:steam:abstractions}

The goal of \toolname{} is to support any operational technique (including sustainability strategies) with minimal implementation effort, regardless of which other techniques are enabled. To achieve this, we use a component graph, inspired by reactive programming~\cite{DBLP:journals/csur/BainomugishaCCMM13} at the core of our simulator. This architecture is inherently composable and loosely coupled, enabling the easy integration of new techniques and models.


We define a \textit{component} as any object in the simulator. Components can represent physical datacenter entities, such as CPUs and GPUs, or abstract objects, like resource managers and fault injectors. 
Directed edges connect components, defining supplier-consumer relationships.
For example, a CPU consumes the power supplied by a Power Supply Unit (PSU). 

Simulation is driven by \emph{events} that change the state of components. An example of an event is a task changing its CPU requirements. Depending on the event, connected nodes might also be affected. When a task updates its CPU requirements, the connected CPU will change its power requirements according to its power model. This creates a cascading effect that could potentially traverse the whole component graph.

\begin{figure}[t]
    \centering
    \includegraphics[width=\linewidth]{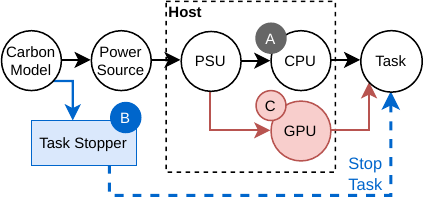}
    \caption{Example component graph depicting events connecting two new components: task stopper and GPU.}
    \label{fig:component-graph}
    \vspace*{-0.35cm}
\end{figure}

The component graph approach of \toolname{} ensures that adding new features only requires changes to directly connected components. This promotes loose-coupling, ensuring that unrelated components remain independent, and allows individual components to be activated or disabled.


\Cref{fig:component-graph} illustrates this composability. 
We start with a simple component graph~\encircle{A} connecting a power source to the PSU, the PSU to the CPU, and the CPU to the workload task. The carbon model provides the power source with the current carbon intensity. 
We extend the component graph by adding a task stopper~\encircleblue{B}. The task stopper monitors the carbon intensity of the power source and stops tasks when the carbon intensity exceeds a threshold to reduce the datacenter's operational carbon. 
Next, we add a GPU~\encirclered{C} to the host. A GPU is connected to only two nodes; it demands power from the PSU and provides computation for the task. The GPU is inserted into the component graph and connected to the PSU and the task. Besides the directly connected nodes, no other nodes or edges have to be updated.

We use this composability to implement three sustainability techniques: hardware scaling, batteries, and temporal shifting. 
This required us to implement components to start and stop tasks based on carbon intensity, batteries, and a battery resource manager. All these techniques are independent and can be activated by the user with a simple configuration change. This flexibility allows \toolname{} to analyze techniques independently and in conjunction with each other or other simulator features, including fault injection, checkpointing, restarts, and advanced scheduling mechanisms.


\section{Experimental Setup}\label{sec:setup}
The following sections discuss the setup used for all experiments and evaluations. To ensure generalizability, we evaluate sustainability techniques across diverse workloads, hardware configurations, and carbon-intensity traces. All techniques, costs, and parameters are configurable in \toolname{}.


\subsection{Workload and Datacenter setup}
\subsubsection{Workload Traces}\label{subsubsec:setup:workload_traces}
The characteristics of a workload can significantly influence the effectiveness of sustainability techniques. To capture this, we experiment with three publicly available workloads from different datacenter types: Surf LISA~\cite{DBLP:journals/fgcs/VersluisCGLPCUI23}, CINECA Marconi M100~\cite{borghesi2023m100}, and Google Borg~\cite{tirmazi2020borg}. The workloads differ in duration, task frequency, and task execution time (see \Cref{tab:setup:workloads}). The Lisa workload is 5 months long and was collected in 2019 by SURF, an IT cooperative of Dutch education and research institutions\footnote{https://www.surf.nl/en}. The Marconi M100 workload is collected by CINECA using their EXAMON monitoring framework~\cite{borghesi2023m100}. Over 90\% of the tasks in the Marconi trace utilize GPUs. We use a subset of the dataset, recorded in October 2022. Finally, we use a large-scale industrial workload based on the Google Borg infrastructure~\cite{tirmazi2020borg}.
We use a subset of the dataset containing all tasks executed on cell-a located in New York\footnote{\url{https://console.cloud.google.com/storage/browser/clusterdata_2019_a;tab=objects?authuser=0&invt=AbuSoA}}.
Tasks can not run on more than one server, but multiple tasks can share the same server. Tasks can be CPU and GPU-based. 


\begin{figure*}[t]
    \centering
    \includegraphics[width=\linewidth]{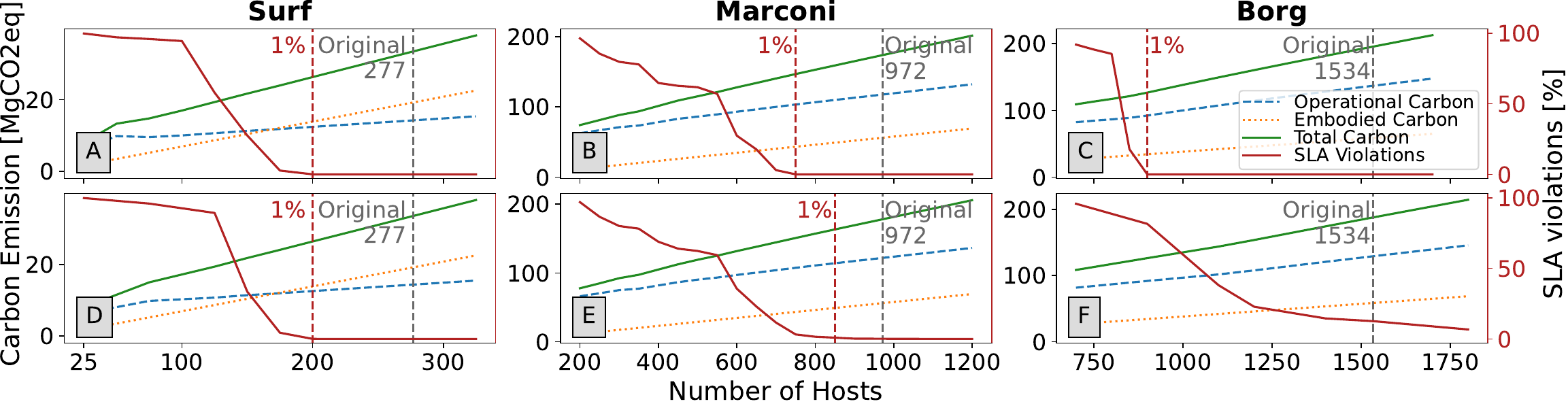}
    \caption{Impact of horizontal scaling on total carbon emissions (operational + embodied) and performance, quantified by SLA violations (tasks not scheduled within 24 hours of submission). For all metrics, lower is better. Gray and red lines indicate the original and required scale for \textless1\% SLA violations. Bottom row shows the impact when the datacenter is exposed to failures.}
    \label{fig:horizontal_scaling}
\end{figure*}

\subsubsection{Datacenter setup}\label{subsubsec:setup:available_hardware}
The hardware in a datacenter can significantly impact its performance and sustainability. Each experiment has a corresponding datacenter setup, reflecting the hardware (CPU, GPU, RAM) used when collecting the traces (see \Cref{tab:setup:hardware}). Unless stated otherwise, workloads are run on these datacenters.

\subsection{Sustainbility techniques}
\subsubsection{Batteries}\label{subsubsec:setup:batteries}
Batteries are used to reduce a datacenter's carbon emissions (see \Cref{subsec:techniques}). The policy we use to determine when to charge and discharge is similar to previous works~\cite{li2017balancing, sheme2018battery} and is based on a carbon threshold: if the current carbon intensity is below the threshold, the battery is charged; if it is above the threshold, the battery is discharged. The threshold is equal to the rolling mean of the past week's carbon intensity. As an optimization, batteries wait until the carbon intensity stops decreasing before charging.

The \textit{battery capacity} can significantly impact its effectiveness: larger battery capacities allow the datacenter to run longer on battery power, but will also increase the embodied carbon cost. We evaluate multiple capacities and report the configuration achieving the highest total emissions reduction.

We assume that the \textit{charging speed} of batteries increases linearly with their capacity due to parallel charging. The charging speed of batteries scales with a factor of 3 kW/kWh, similar to the charging speed of a Tesla Model 3 using a DC charger\footnote{\url{https://evbox.com/en/electric-cars/tesla/tesla-model-3}}. In \Cref{subsub:BAT:capacity_charging}, we further discuss the impact of battery capacity and charging speed. 

\subsubsection{Temporal Shifting}\label{subsubsec:setup:temporal_shifting}
Temporal shifting is the process of delaying tasks to better align with the availability of low-carbon energy. Many past works have used "oracle" models to determine when to schedule a task optimally. In this work, we use a task-scheduling policy similar to~\cite{bostandoost2024data}. A task is scheduled when the current carbon intensity is below the carbon threshold. Otherwise, they are delayed. The carbon threshold is dynamic and is set to the 35th percentile of next week's carbon forecast. All tasks have a maximum delay time of 24 hours and are interruptable. When the maximum delay is reached, a regular FIFO scheduler is used.


\begin{table}[t]
    \centering
    \caption{Workload Traces used in the evaluation. NoT: number of tasks. ATD: Average task duration in HH:MM:ss.}
    \begin{tabular}{c|c|r|r|c}\toprule
    \textbf{Workload} & \textbf{Start} & \textbf{Runtime} & \textbf{NoT} & \textbf{ATD}\\\hline
        Surf~\cite{DBLP:journals/fgcs/VersluisCGLPCUI23} & 7-2022 & 124 days & 194,917 & 01:49:38 \\
        Marconi~\cite{borghesi2023m100} & 9-2022 & 30 days & 73,882 & 06:20:12 \\
        Borg~\cite{tirmazi2020borg} & 5-2019 & 31 days & 14,867,803 & 02:01:51 \\\bottomrule
    \end{tabular}
    \label{tab:setup:workloads}
\end{table}

\begin{table}[t]  
    \small
    \centering
    \caption{Default hardware setup per workload. NoH: Number of Hosts, EC: Host Embodied carbon [kgCO2-eq].}
    \resizebox{\columnwidth}{!}{
    \begin{tabular}{c|r|c|r}\toprule
    \textbf{Source} & \textbf{NoH} & \textbf{Host resources} & \textbf{EC} \\\hline
        Surf~\cite{DBLP:journals/fgcs/VersluisCGLPCUI23} & 277 & \makecell{ 128~GB RAM, 16-core\\ 2.1GHz Intel Xeon Silver 4110} & 1,022 \\
        \hline Marconi~\cite{borghesi2023m100} & 972 & \makecell{196~GB RAM, 2x24-core\\ 2.1GHz Intel Xeon 8160,\\
        4xNVIDIA VoltaV100 16GB} & 3,542\\
        \hline
        Borg~\cite{tirmazi2020borg} & 1,534 &  Obfuscated (see \Cref{subsec:setup:validation}) & 2,250\\\bottomrule
    \end{tabular}
    }
    \label{tab:setup:hardware}
    
\end{table}

\subsection{Carbon Emissions}
\subsubsection{Operational Carbon}\label{subsubsec:setup:operational_carbon}
Operational carbon, emitted when using energy, is determined by multiplying energy usage by the power source's carbon intensity (amount of carbon emissions per unit of energy). The power draw of CPUs and GPUs is modeled using a square root and linear model, respectively, following approaches adopted in prior state-of-the-art work~\cite{exadigit}. Historical carbon intensity traces collected from the ElectricityMaps portal \cite{ElectricityMaps2025} are used to determine the carbon intensity during a workload. 
To capture a broad range of scenarios, we use 158 carbon traces from different regions collected in 2021-2024, which display diverse carbon intensity profiles and coefficients of variation. Some experiments are only run in a single region. In those cases, we use the datacenter's region. These are the Netherlands, Italy-North, and New York ISO for Surf, Marconi, and Borg, respectively. 

\subsubsection{Embodied Carbon}\label{subsubsec:setup:embodied_carbon}
The embodied carbon of a workload is estimated by multiplying the fraction of the hardware's total lifetime used by the workload's runtime with the total embodied cost~\cite{Gupta2022act}. In this work, the embodied cost of a datacenter is based on the host and battery.

For \textit{hosts}, we assumed a 5-years lifespan. A host's embodied carbon cost is based on its components. Unfortunately, manufacturers rarely disclose their components' embodied carbon. Therefore, we estimated hosts' embodied carbon using EC2 instance equivalents from a dataset collected by Teads Engineering~\cite{buildingAWS}. 
We match Surf and Marconi hardware to the \textit{a1.4xlarge} and \textit{p3.8xlarge} instances, resulting in embodied carbon costs of 1,022 and 3,542 kgCO2-eq, respectively.
For Borg, we do not have hardware specifics (see \Cref{subsec:setup:validation}). Instead, we estimate the embodied carbon to contribute 40\% of the total carbon emissions, a common distribution for modern datacenters~\cite{Gupta2022act}. For Borg, this is 2,250 kg CO2-eq. 

For \textit{batteries}, we assumed a 10-years lifespan. The embodied carbon of a battery depends on its manufacturing process and location, typically ranging from 30 to 500 kg CO2-eq/kWh~\cite{dale2018battery}. For our experiments, we use an embodied carbon cost of 100 kg CO2-eq/kWh.


\section{Analysis of Sustainability Techniques}\label{sec:exp_individual}
In the following sections, we demonstrate \toolname{} capabilities by investigating the impact of horizontal scaling, batteries, and temporal shifting, on datacenter carbon footprint. 

\subsection{Horizontal Scaling}\label{subsec:horizontal_scaling}
We use \toolname{} to investigate the effects of horizontally down-scaling datacenters on their performance and sustainability. We measure performance using \textit{Service Level Agreement} (SLA) violations: a task meets the SLA if it is completed within 24 hours of its expected completion time. We define a datacenter service as acceptable if it has fewer than 1\% SLA violations. We aim to find the smallest datacenter configuration that still meets this threshold. Since horizontal scaling is independent of carbon intensity, we run experiments in a single region determined by the datacenter's location.


\begin{figure*}
    \centering
    \begin{minipage}[t]{0.34\textwidth}
        \centering
        \includegraphics[width=\textwidth]{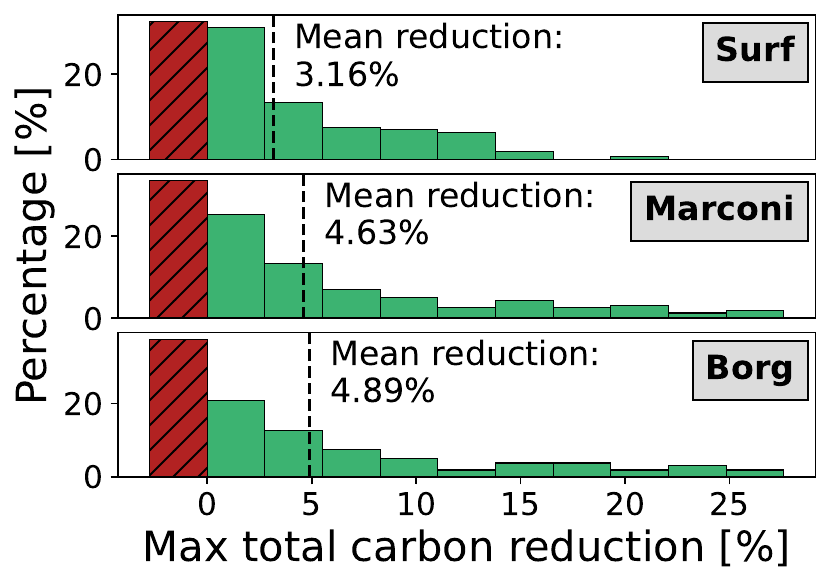}
        \begin{minipage}[h]{0.95\textwidth}
            \centering
            \caption{Total carbon emission reduction using batteries in 158 carbon regions on three workloads. Red bars mark all regions where batteries \textit{increased} emission.}\label{fig:battery_impact}
        \end{minipage}
    \end{minipage}
        \begin{minipage}[t]{0.32\textwidth}
        \centering
        \includegraphics[width=\textwidth]{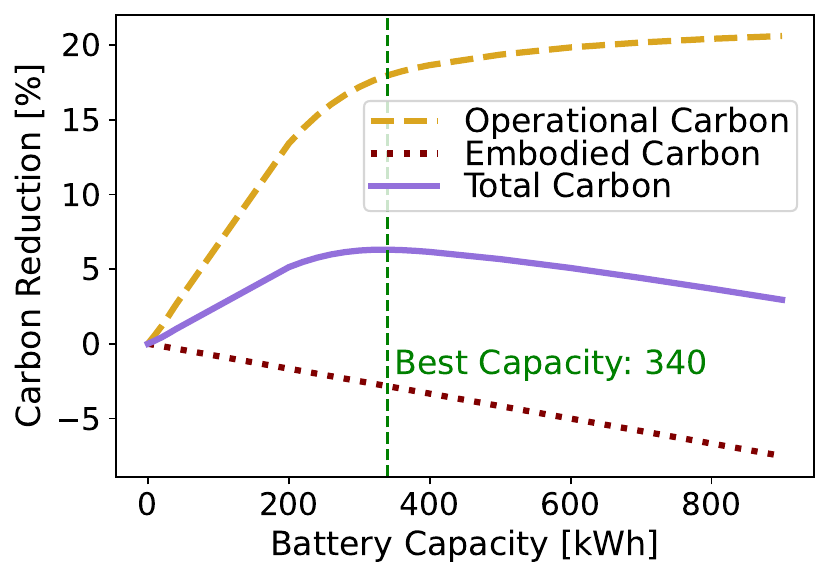}
        \begin{minipage}[h]{0.95\textwidth}
            \centering
            \caption{The impact of battery capacity on the effectiveness of using batteries to reduce carbon emissions.}
            \label{fig:battery_capacity}
        \end{minipage}
    \end{minipage}
    \begin{minipage}[t]{0.32\textwidth}
        \centering
        \includegraphics[width=\textwidth]{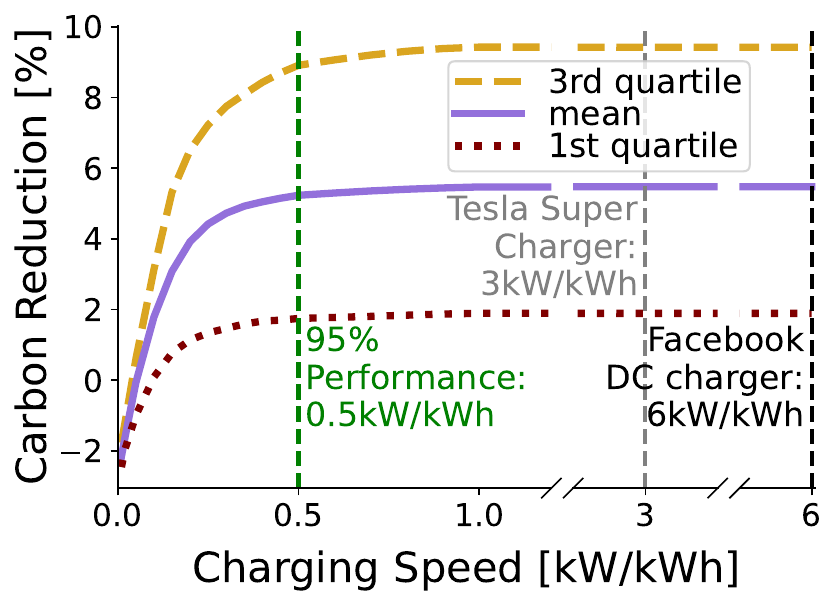}
        \begin{minipage}[h]{0.95\textwidth}
            \centering
            \caption{The impact of battery charging speed) on the effectiveness of using batteries to reduce carbon emissions.}
            \label{fig:battery_charging}
        \end{minipage}
    \end{minipage}
\end{figure*}

\subsubsection{Performance}\label{subsubsec:HS:performance}
\Cref{fig:horizontal_scaling} shows the impact of horizontal scaling on system performance. Plots A, B, and C show the effects on systems without failures. We define a datacenter as \textit{underprovisioned} if it has more than 1\% SLA violations and \textit{overprovisioned} otherwise.

\textit{Under-provisioning:} When a datacenter does not have sufficient hosts to run incoming tasks, they are delayed until a host becomes available. Since tasks arrive continuously, delays accumulate, leading to large SLA violations that approach 100\%. Adding more machines does not always reduce SLA violations significantly. For example, scaling the Surf datacenter from 25 to 100 machines reduces SLA violations by only 5\% (see \Cref{fig:horizontal_scaling}A). Even with 100 machines, the tasks still accumulate. Interestingly, the 4x increase in scale does not significantly increase operational carbon emissions, as the increased energy usage is offset by a shorter runtime.

\textit{Over-provisioning:} Datacenters must have sufficient hosts to absorb workload peaks and avoid SLA violations. We find that 200, 750, and 900 machines are sufficient for the Surf, Marconi, and Borg workloads, respectively. Adding machines yields minimal SLA improvement but increases carbon emissions due to higher idle power draw and embodied carbon.
Reducing the number of machines results in an average carbon reduction of 14\%, 12\%, and 35\%, respectively.

\subsubsection{Failures}\label{subsubsec:HS:failures}
Failures (e.g., hardware or task failures) can severely disrupt service quality. To mitigate these risks, operators typically over-provision resources. We use \toolname{} to simulate the impact of failures of the operations of a datacenter. For this analysis, we configured \toolname{} to inject machine failures using a trace from the Cloud Uptime Archive, derived from incident reports for Facebook Messenger~\cite{CUA}. To lessen the impact of failures, we use checkpointing, which periodically saves task snapshots. In case of a failure, the task resumes from its latest snapshot rather than starting over. We use a checkpointing frequency of 1 hour similar to~\cite{CUA}.

Plots D, E, and F in \Cref{fig:horizontal_scaling} show the impact of horizontal scaling on a datacenter subjected to failures for the Surf, Marconi, and Borg workloads. The Surf workload is minimally affected, with the scale required for \textless 1\% SLA violations staying at 200. Marconi is moderately impacted: the number of hosts needed to get below 1\% SLA violations is increased from 750 to 850. This reduces the carbon reduction for Marconi from 12\% to 5\%. The Borg workload is the most affected: with failures, it is not able to achieve \textless 1\%, with the default scale having 12.5\% SLA violations. This could mean that Borg has a lower failure rate than we used or uses a different SLA.  

\begin{keytakeawaybox}
\defkeytakeaway{Horizontal_Scaling} Down-scaling over-provisioned datacenters can reduce total carbon-emissions by up to 35\% without impacting service quality. \textit{However}, when accounting for operational phenomena such as failures, the achievable reduction falls to 14\%. \toolname{} can accurately quantify these trade-offs to help optimize designs and operations.
\end{keytakeawaybox}


\subsection{Batteries}\label{subsec:BAT}

In this section, we use \toolname{} to assess the impact of using batteries on a datacenter's carbon footprint. 


\subsubsection{Battery Performance}\label{subsub:BAT:performance}
\Cref{fig:battery_impact} shows the total carbon reduction achieved using batteries in 158 carbon regions 
while running the Surf, Marconi, and Borg workloads on their respective datacenters. On average, batteries reduced total carbon by 3.16\%, 4.63\%, and 4.89\% for the three workloads, respectively. However, there is a significant difference in battery effectiveness across the carbon regions. Using batteries significantly decreases carbon emissions ($\geq$ 5\%) in 24\%, 29\%, and 33\% of the regions for the Surf, Marconi, and Borg workloads, respectively, with the highest reduction being 21\%, 27\%, 28\%. However, in 32\%, 34\%, and 37\% of the regions, carbon emissions actually increased, caused by the additional embodied carbon outweighing the reduced operational carbon.


\begin{keytakeawaybox}
\defkeytakeaway{BAT_effectiveness} Batteries' effectiveness depends on the datacenter, workload, and energy source.  \toolname{} shows that batteries reduce carbon emissions significantly~($>5$\%) in 136 out of the 474 scenarios we evaluated. \textit{However}, when the variation in regional carbon intensity is low, batteries can actually increase total carbon emissions.
\end{keytakeawaybox}

\subsubsection{Battery Capacity And Charging Speed}\label{subsub:BAT:capacity_charging}
\begin{figure}[t]
    \centering
    \includegraphics[width=\linewidth]{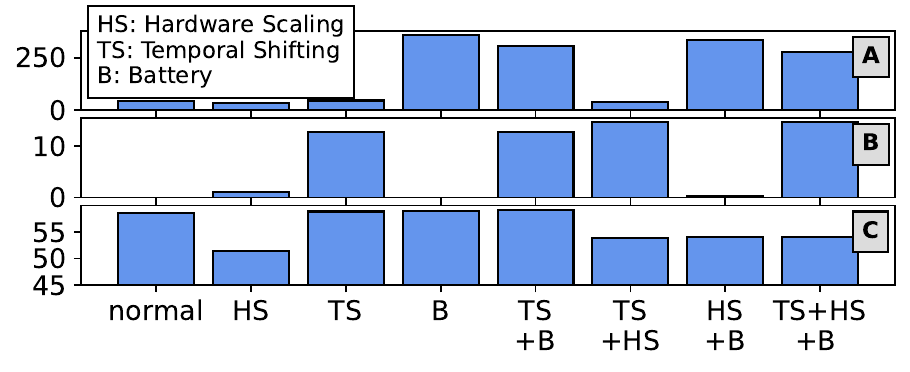}
    \caption{Peak power draw\,[kW] (A), mean task delay[h] (B), and total energy usage\,[MWh] (C) for the Surf workload.} 
    \label{fig:technique_impact}
\end{figure}

Using large fast-charging batteries can be expensive and introduce large power draw spikes, which can stress the power delivery infrastructure. \Cref{fig:technique_impact}A shows that using batteries can increase the peak power draw by 8x when running Surf, with similar results observed for the other workloads.

\textit{Capacity:} \Cref{fig:battery_capacity} shows the impact of battery capacity on carbon emission reductions for the Surf workload. Increasing battery capacity reduces the operational carbon, but with diminishing returns. Batteries reduce operational carbon by providing energy when carbon intensity is high. If a battery can handle most periods of high carbon, increasing its size has little to no impact. 
In contrast, the embodied carbon of a battery increases linearly with its capacity. This creates a sweet spot in which operational carbon reduction exceeds the added embodied carbon, which can be identified using \toolname{}. 

\textit{Charging Speed:} \Cref{fig:battery_charging} shows the impact of battery charging speed on carbon emission reductions for the Surf workload. Increasing charging speed increases effectiveness, but again,  with diminishing returns.
We observed that at a charging rate of 0.5 kW/kWh, 95\% of the mean performance has already been reached. This is 6 times smaller than the charging rate of a Tesla Model 3 DC charger. 

\begin{keytakeawaybox}
\defkeytakeaway{BAT_charging} Optimally using batteries to reduce carbon emissions requires fast-charging batteries with high capacity, which are expensive and can increase the max power-draw up to 8x. \textit{However,} charging speed can be reduced by 50\% while retaining 95\% of the expected carbon reduction.
\end{keytakeawaybox}



\begin{figure}[t]
    \centering
    \includegraphics[width=.9\linewidth]{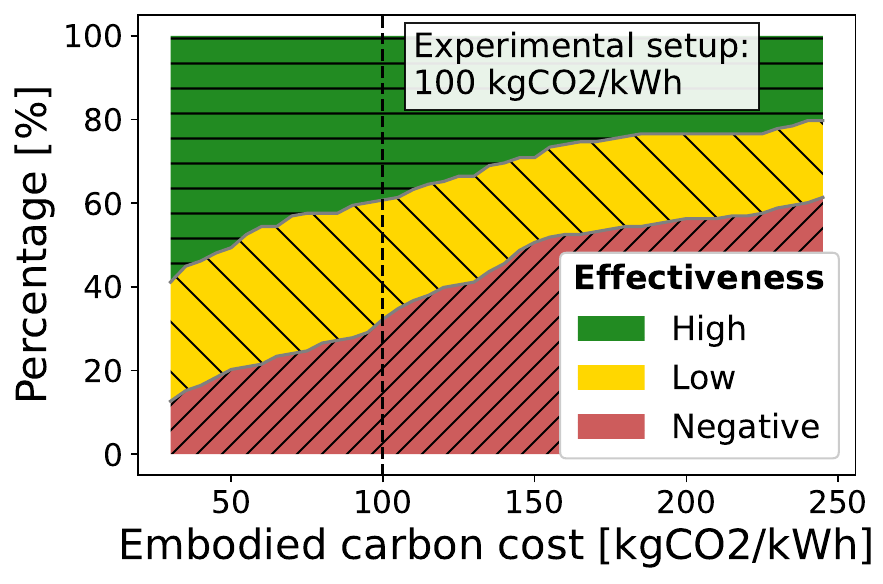}
    \caption{Percentage of regions where the effectiveness of batteries to reduce total carbon emissions is high (\textgreater5\%), low (0\%-5\%), or negative (\textless0\%). The black line indicates the embodied cost (100 kgCO2/kWh) used in other experiments.} 
    \label{fig:battery_embodied}
\end{figure}

\subsubsection{Battery Embodied Carbon}\label{subsub:bat:embodied_carbon}

A major downside of batteries is the carbon cost associated with manufacturing. The embodied carbon of a battery depends on how and where it is manufactured~\cite{dale2018battery}. \Cref{fig:battery_embodied} shows the effectiveness of batteries for embodied carbon costs between 30 and 250 kgCO2-eq/kWh when running Surf. As expected, reducing the carbon cost improves the impact of using batteries. At 30 kgCO2/kWh, the number of regions in which batteries can reduce overall carbon by \textgreater5\% increases from 39\% to 59\%. However, for 13\% of the regions, batteries still increase overall carbon emissions. In these regions, operational carbon reduction is minimal, likely due to a low coefficient of variation in carbon intensity. Batteries are not viable for these regions with the current technology. 



\subsection{Temporal Shifting}\label{subsec:temporal_shifting}
\begin{figure}[t]
    \centering
    \includegraphics[width=\linewidth]{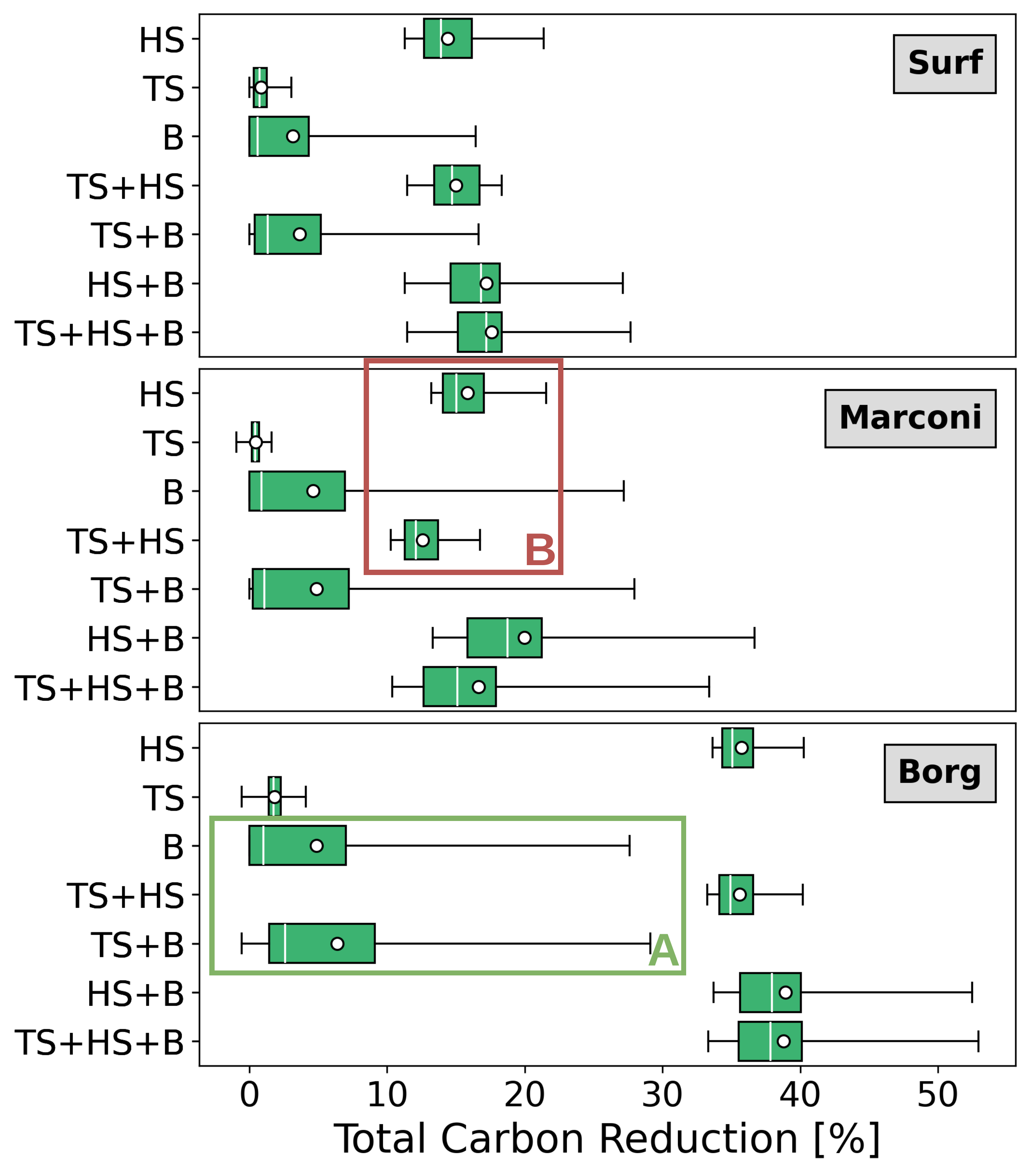}
    \caption{Carbon emissions reduction of temporal shifting (TS), horizontal scaling (HS), and batteries (B) individually and in combination for three workloads in 158 regions. The white dot and line indicate the mean and median. Whiskers represent outliers that are at least 3 times the mean.}
    \label{fig:combined_techniques}
\end{figure}

Temporal shifting reschedules tasks to periods with low-carbon intensity (see \Cref{subsubsec:setup:temporal_shifting}).
\Cref{fig:combined_techniques} depicts the distribution of operational carbon reduction achieved using temporal shifting in 158 carbon traces while running the Surf, Marconi, and Borg workloads. When using temporal shifting, operational carbon is reduced on average by 2.06\%, 0.74\%, and 2.85\% for the three workloads. The effectiveness of temporal shifting varies significantly across workloads and carbon regions, achieving a \textgreater5\% reduction in the best scenarios, whereas it has little impact in others. Using temporal shifting introduces an average delay of 14 hours on the Surf trace (see \Cref{fig:technique_impact}B), which is considerably less than the allowed 24 hours. The carbon reduction achieved by temporal shifting in this work is significantly lower than that reported in previous work, which reported reductions up to 40\%~\cite{sukprasert2024limitations}. We indicate the following reasons for this significant difference: \textit{i) Policy:} Most previous work used oracles that could determine the optimal time to execute a task~\cite{sukprasert2024limitations}. 
However, these models are difficult (or even impossible) to use in real systems. On the contrary, the policy used in this work can easily be adapted to most datacenters. More advanced scheduling policies could improve the effectiveness, but would still not match the oracle models.
\textit{ii) Idle Hosts:} Previous work evaluated the impact of temporal shifting by only looking at the tasks~\cite{sukprasert2024limitations, bostandoost2024data} (see \Cref{sec:motivation}). This ignores the idle power draw of hosts. \textit{iii) Compounding Tasks:} Previous work did not account for tasks stacking when multiple tasks are delayed to the same timestamp (see \Cref{sec:motivation}). STEAM automatically accounts for such operational phenomena, quantifying their impact on performance and carbon emissions.

\begin{keytakeawaybox}
\defkeytakeaway{Shifting} Temporal shifting reduces carbon emissions by over 2\% in 210 of the 474 considered scenarios. These results are significantly lower than previously reported at the state of the art; \toolname{} considers the impact of complex dynamics such as compounding tasks. 
\end{keytakeawaybox}


\section{Analysis of Combined Sustainability Techniques}\label{sec:exp_combined}
Prior works have primarily focused on using single sustainability techniques. In contrast, the composable design of \toolname{} (\Cref{sec:steam}) enables assessing the impact of combining multiple techniques. In the following, we explore the effectiveness of compositing the three sustainability techniques discussed so far. 
\ifarXiv
\else
An extended version of this paper, including additional results, can be found at \url{https://arxiv.org/abs/2603.12381}.
\fi

\Cref{fig:combined_techniques} shows the distribution of total carbon reduction achieved by individual and combined sustainability techniques.
As expected, in some cases, combining sustainability techniques can reduce carbon emissions more than when used individually. For example, combining batteries with temporal shifting when running Borg increases the average reduction from 4.9\% (batteries alone) to 6.4\% (\Cref{fig:combined_techniques}, box A). This is slightly less than the sum of the individual reductions.
However, there are also combinations of techniques, workloads, and locations that have no impact on effectiveness or even reduce it. For example, on Marconi, combining horizontal scaling and temporal shifting lowers the reduction from 15\% to 12.5\% compared to when using only horizontal scaling (\Cref{fig:combined_techniques}, box B).

\begin{keytakeawaybox}
\defkeytakeaway{Combined_res}
\toolname{} helps systematic analysis of how sustainability techniques interact.
Some combinations, such as batteries and horizontal scaling, are compatible, achieving an average carbon reduction close to the sum of their individual effects. 
\textit{However,} combining other techniques, such as temporal shifting and horizontal scaling, can lead to lower average carbon-reduction when combined. The effectiveness of any combination also depends on the workload characteristics, the datacenter, and its location.
\end{keytakeawaybox}

\begin{figure}[t]
    \centering
    \includegraphics[width=\linewidth]{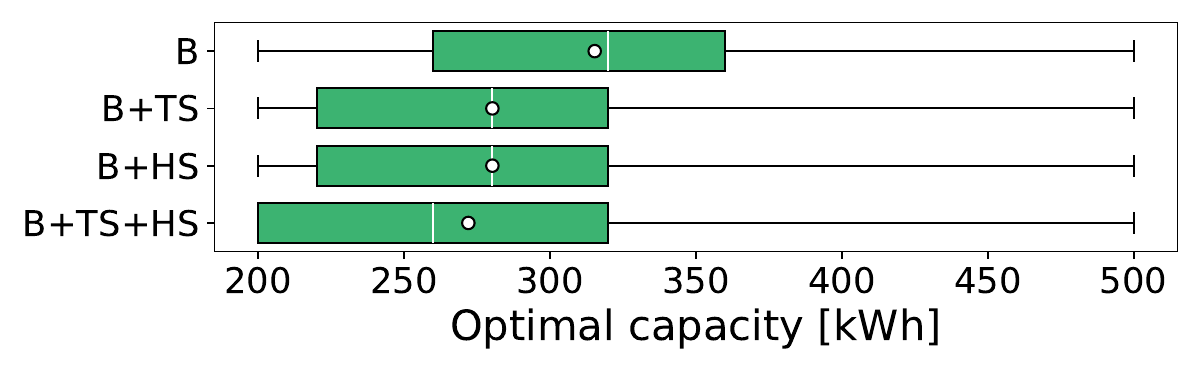}
    \caption{The optimal battery size in the range [200, 500] kWh for the combinations of sustainability techniques across 158 regions. The white dot and line indicate the mean and median. Whiskers represent outliers that are at least 3 times the mean.}
    \label{fig:optimal_battery}
\end{figure}

Although combining sustainability may seem straightforward, understanding its implications is not. Further investigation shows that combining techniques optimally could require different setups than when used individually. For example, the optimal battery size decreases when combined with other techniques. \Cref{fig:optimal_battery} shows the optimal battery size when running the Surf workload across different regions. The average optimal battery size decreases from 315 kWh to 272 kWh when all techniques are combined.

\begin{keytakeawaybox}
\defkeytakeaway{Combined_Bat_size} Combining techniques can impact their optimal setup. For instance, the optimal battery size is reduced when combined with temporal shifting.
\end{keytakeawaybox}

The findings in this work show that the simulation-based approach of \toolname{} and its composable design are necessary to accurately assess the impact of sustainability techniques, both individually and in combination, while capturing their interactions with datacenter operational phenomena.

\section{Calibration, Validation, and Performance}\label{subsec:setup:validation}

\toolname{} uses power models to estimate the power draw of hardware components such as the CPU and GPU, based on parameters such as max and idle power draw. These models are calibrated through a two-step process. \textit{1)} We \textit{select initial values} for Surf and Marconi based on information provided by manufacturers, indicating expected power draw.  
\textit{2)} 
We further \textit{calibrate} the Surf and Marconi datacenters on a subset of the existing data (1 month and 1 day, respectively)
The accuracy of \toolname{} is evaluated by comparing the estimated power draw against real-world measurements. After calibration, \toolname{} achieves a Mean Average Percentage Error of 5.5\% and 3.4\% for Surf and Marconi, respectively. These results highlight \toolname{} ability to realistically estimate energy usage, making it a reliable tool for sustainability analysis for both CPU-(Surf) and GPU-based (Marconi) workloads. 

Calibrating Borg was less straightforward, as the trace is highly obfuscated to protect confidential or proprietary information~\cite{reiss2012obfuscatory}. Critical information, such as hardware specification, resource utilization, or power usage, is hashed or converted to a normalized metric. However, not everything is unknown. First, we know the number of nodes is 1534. Second, resources have an average utilization of 60\%~\cite{tirmazi2020borg}. Finally, the trace we use is based on a single PDU, with a max power draw of 2 MW~\cite {sakalkar2020data}. We convert the normalized utilization number to match the average resource utilization of 60\% and the max power draw of 2MW. The power models are calibrated to match these derived values. Given this calibration approach, the simulation is consistent with the trace under the assumed conditions but cannot be independently validated.

The simulation time of \toolname{} depends on the workload size, the datacenter hardware, and the techniques used. As a reference, a single simulation of Surf, Marconi, and Borg took 1, 2, and 10 minutes on average. Simulations are independent and can be executed in parallel. All experiments were executed on a 20-node cluster, each equipped with a single 24-core 2.8 GHz AMD EPYC-2 (Rome) 7402P CPU and 128 GB of RAM. 
In this paper, we simulated over 2,787 years of datacenter operations using just 60 hours of actual compute time. During these simulations, 83 billion tasks were executed, corresponding to 170 billion simulated compute hours.

\section{Related Work}\label{subsec:related_work}
In past years, many techniques have been proposed to improve the carbon footprint of a datacenter. This work quantifies the impact of three such techniques:  horizontal scaling~\cite{lin2012dynamic,  uddin2010server}, batteries~\cite{acun2023carbon, sheme2018battery, li2017balancing}, and load shifting~\cite{acun2023carbon, sukprasert2024limitations}.

Lin et al. propose dynamic "right-sizing" in which machines are turned off during low-load periods, showing that this can reduce energy usage by at least 15\%~\cite{lin2012dynamic}. Uddin et al. show that dynamic server consolidation can reduce energy usage by 50\%~\cite{uddin2010server}, but do not quantify the impact on carbon emission. Sukprasert et al. use analytical models to determine the limitations of temporal shifting~\cite{sukprasert2024limitations}. Sheme et al. use a finite-state machine to show that batteries can improve green energy coverage~\cite{sheme2018battery}. However, their work is limited to three countries and a single datacenter with constant power.
%


Carbon Explorer uses analytical modeling to assess the impact of sustainability techniques on achieving NetZero carbon emissions ~\cite{acun2023carbon}. However, it does not model hosts or tasks, but only the total datacenter energy usage. This makes it hard to understand task-level effects such as task stacking or SLA violations (see \Cref{sec:motivation}). Additionally, the tool does not specify how a datacenter should implement temporal shifting. Instead, it simply shifts the total power draw to highlight the potential carbon reductions. Due to this coarse granularity, only the analysis in \Cref{fig:battery_impact} can be replicated by Carbon Explorer without requiring significant modifications to the tool design and implementation.


We are not the first to use simulation to explore datacenters and their carbon footprint~\cite{song2022versatility}. 
\toolname{} builds on and significantly extends OpenDC~\cite{mastenbroek2021opendc}, an open-source datacenter simulator used in prior works on sustainability~\cite{niewenhuis2024footprinter}. Although inspired by OpenDC's core concepts, \toolname{} required a major architectural redesign to: \textit{i)} enable modularity and composability, crucial for modeling and analyzing multiple sustainability techniques and their interactions; \textit{ii)} introduce sustainability techniques and carbon footprint modeling; and \textit{iii)} support GPU-enabled datacenters. This flexibility and novel features are core contributions of this work, allowing service providers and sustainability researchers to explore how different techniques and their interactions impact modern datacenters’ design and operations.
%
Popular alternatives to OpenDC are CloudSim~\cite{calheiros2011cloudsim}, and SimGrid~\cite{SimGrid}, two datacenter simulators that have spawned many more specialized simulators. Using these specialized simulators in combination is difficult, limiting their use. In contrast, \toolname{} allows users to mix and match all implemented components and techniques.

\section{Threads to Validity}\label{sec:Threads}
We recognize and acknowledge that our analysis is not immune to threats to validity, which we outline below to ensure transparency and guide future work.

The \textit{policies used} for batteries and temporal shifting are practical and easy for most datacenter operators to implement. Although this improves applicability, we acknowledge that more advanced strategies may yield better results. OpenDC~\toolname{} enables convenient exploration of different policies, offering a convenient framework for future research. When using horizontal scaling, we used SLA violations to determine the optimal scale. \toolname{} allows users to define different SLAs (e.g., with stricter completion-time constraints), which could lead to different optimal scales and carbon reductions. 

We computed \textit{operational carbon} using historical carbon traces. However, large datacenters can have special contracts with energy providers, resulting in a different energy mix. \toolname{} can be used to analyze the impact of such contracts. 

In this work, energy usage is determined by the computational load of CPUs and GPUs. We acknowledge that other components of datacenters might significantly contribute to carbon costs, such as storage, networking, and cooling costs. Future work can easily incorporate models for these components, further expanding \toolname{} capabilities.

We estimate \textit{embodied carbon} based on prior industry reports~\cite{buildingAWS, Gupta2022act}. Accurate embodied carbon estimation is an active area of research, and different models can yield widely different carbon cost estimates~\cite{bhagavathula2024understanding}. As more accurate cost models or transparent manufacturer reports become available, these can be easily integrated into \toolname{}. 

We used CO2 emissions as a proxy to estimate the sustainability impact of different techniques and design choices. Although CO2 is the most common metric of sustainability, a more comprehensive picture of environmental impact could be obtained by extending \toolname{} to model other aspects, such as water wastage.

\section{Conclusion}\label{sec:conclusion}

We introduced OpenDC~\toolname{}, a trace-based, composable simulation tool to quantify the impact of sustainability techniques on datacenters' design and operations. \toolname{} can provide insights into sustainability aspects, such as carbon emissions, as well as performance, such as scheduling delays or peak power draw.
We used \toolname{} to evaluate three representative sustainability techniques: horizontal scaling, temporal shifting, and batteries. 
We showed that all three techniques can reduce carbon emissions when combined, but their effectiveness depends on the workload, datacenter configuration, and energy sources. They can also introduce trade-offs: for example, our analysis shows that temporal shifting is less effective than previously reported, and that combining techniques creates new operational challenges, such as power spikes, that must be carefully navigated.
%
%
This is the first study to systematically quantify both the individual and combined effects of sustainability strategies across diverse datacenters, workloads, and carbon regions. These results highlight the importance of holistic, quantitative tools such as \toolname{} in guiding sustainable datacenter design and operations.

\changes{We are continuing the development of OpenDC-STEAM, adding more sustainability techniques and additional metrics, such as water consumption and monetary costs. 
Furthermore, we are exploring the use of OpenDC-STEAM's simulation capabilities to support large-scale digital twin projects.  
}


\section*{Acknoweldgement}
This work was partially supported by 
EU MSCA CloudStars (g.a. 101086248) and 
EU Horizon Graph Massivizer (g.a. 101093202). 
This research is partly supported by a National Growth Fund through the Dutch 6G flagship project Future Network Services.

\bibliographystyle{IEEEtran}
\bibliography{references}


\ifarXiv
\clearpage
\appendices

\section{Electricity Maps}\label{sec:app:maps}
In this work, we utilize carbon traces collected from the Electricity Maps platform~\cite{ElectricityMaps2025}. In this section, we discuss the diversity of this dataset. In their paper, Sukprasert et al. provide an extensive analysis of the carbon traces~\cite{sukprasert2024limitations}. We expand on this analysis, increasing the number of regions from 123 to 158, and extending the duration from 2020-2022 to 2021-2024. The traces are collected from the Electricity Datasets page \footnote{\url{https://portal.electricitymaps.com/datasets}} and can be found in the artifacts of this paper. \Cref{fig:app:CI_CoV} shows the average carbon intensity and the daily variability of the carbon intensity of the 158 carbon traces used in this work. We can note a great diversity of both average carbon intensity, ranging between 15 and 860, and daily variance, ranging between near 0 and 0.6.

\begin{figure}[h]
    \centering
    \includegraphics[width=\linewidth]{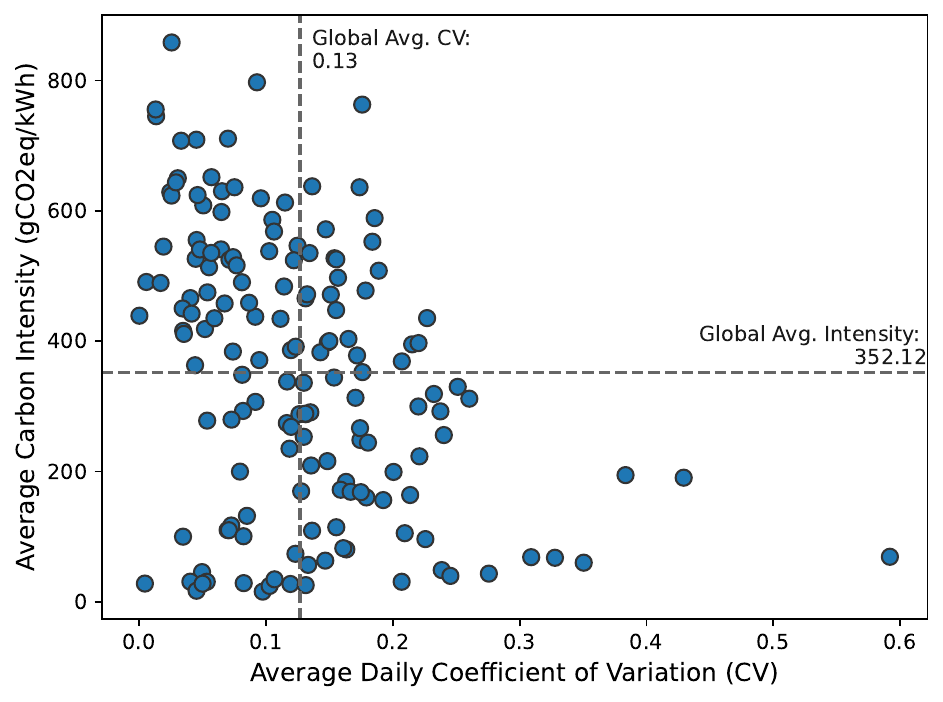}
    \caption{The Average Carbon-Intensity and average daily variability in 2021-2024.}
    \label{fig:app:CI_CoV}
\end{figure}

\begin{figure*}[t]
    \centering
        \begin{minipage}[t]{0.32\textwidth}
        \centering
    \includegraphics[width=\linewidth]{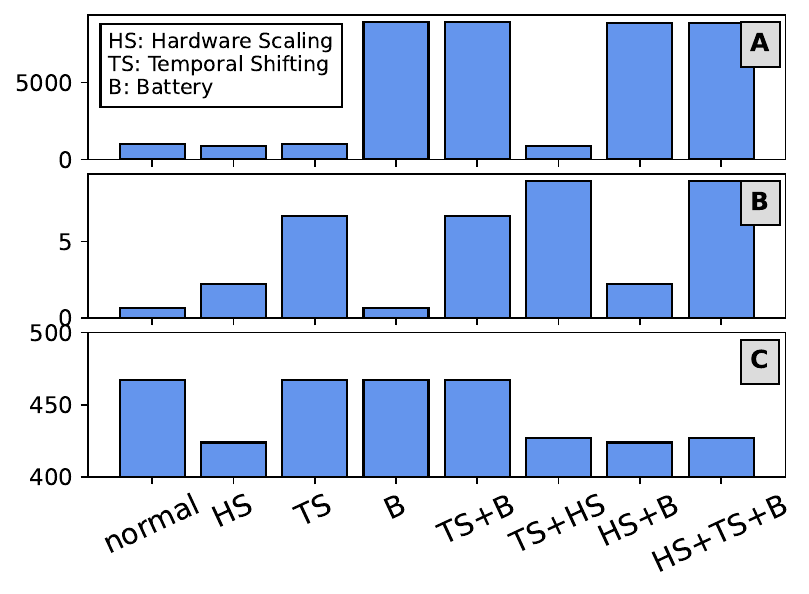}
        \begin{minipage}[h]{0.95\textwidth}
            \centering
            \caption{Max power draw[kW] (A), mean task delay[h] (B), and total energy usage[MWh] (C) for the Marconi workload using different techniques. TS: Temporal Shifting, HS: Horizontal Scaling, B: Batteries.}
            \label{fig:power_spikes_marconi}
        \end{minipage}
    \end{minipage}
    \begin{minipage}[t]{0.32\textwidth}
        \centering
    \includegraphics[width=\linewidth]{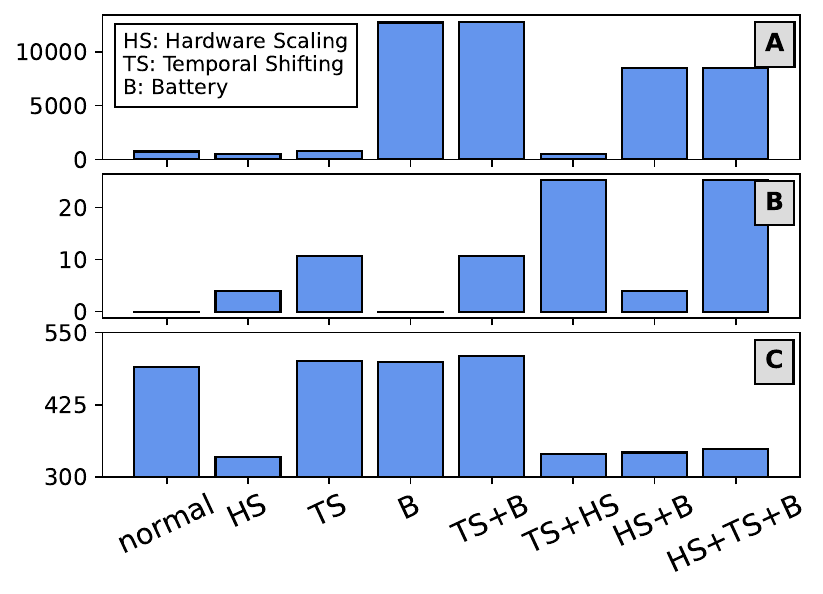}
        \begin{minipage}[h]{0.95\textwidth}
            \centering
            \caption{Max power draw[kW] (A), mean task delay[h] (B), and total energy usage[MWh] (C) for the Borg workload using different techniques. TS: Temporal Shifting, HS: Horizontal Scaling, B: Batteries.}
            \label{fig:power_spikes_borg}
        \end{minipage}
    \end{minipage}
    \begin{minipage}[t]{0.32\textwidth}
        \centering
        \includegraphics[width=\textwidth]{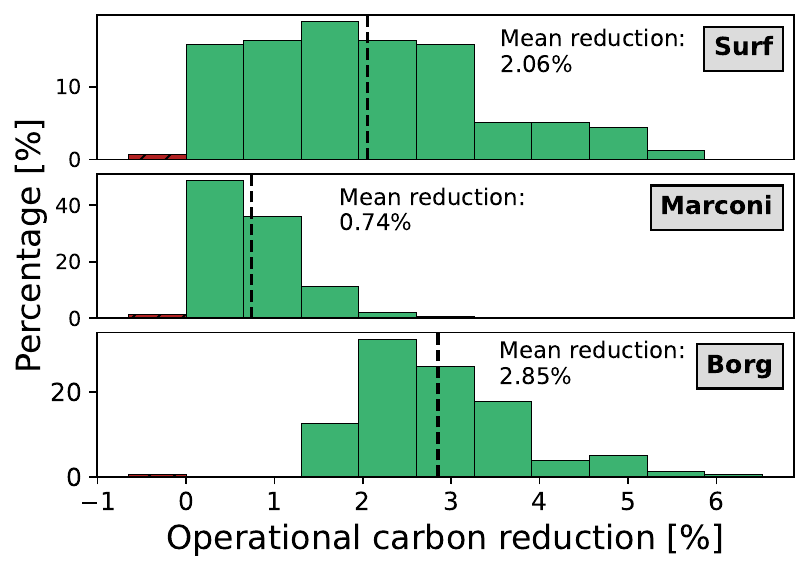}
        \begin{minipage}[h]{0.95\textwidth}
            \centering
            \caption{Total carbon emission reduction using temporal shifting in 158 carbon regions on three workloads.}\label{fig:TS_impact}
        \end{minipage}
    \end{minipage}
\end{figure*}



\section{Additional results}\label{sec:app:additional_results}
\Cref{fig:technique_impact} shows the max power draw, task delay, and total energy usage when using different sustainability techniques on the Surf workload. \Cref{fig:power_spikes_marconi} and \Cref{fig:power_spikes_borg} show the same figure for the Marconi and Borg workloads. The results are very similar, differing only in absolute number due to differences in workload size and topology. This confirms and generalizes our findings in Sections \ref{subsub:BAT:capacity_charging}, \ref{subsec:temporal_shifting}, and \ref{sec:exp_combined}. 

\Cref{fig:combined_techniques} shows the results of all combinations of techniques in a condensed boxplot. Figure \ref{fig:TS_impact}, \ref{fig:TS_BAT_impact}, \ref{fig:HS_BAT_impact}, and X breaks the results down further by showing histograms of the achieved carbon reduction of each combination of techniques.

\begin{figure*}[t]
    \centering
    \begin{minipage}[t]{0.32\textwidth}
        \centering
        \includegraphics[width=\textwidth]{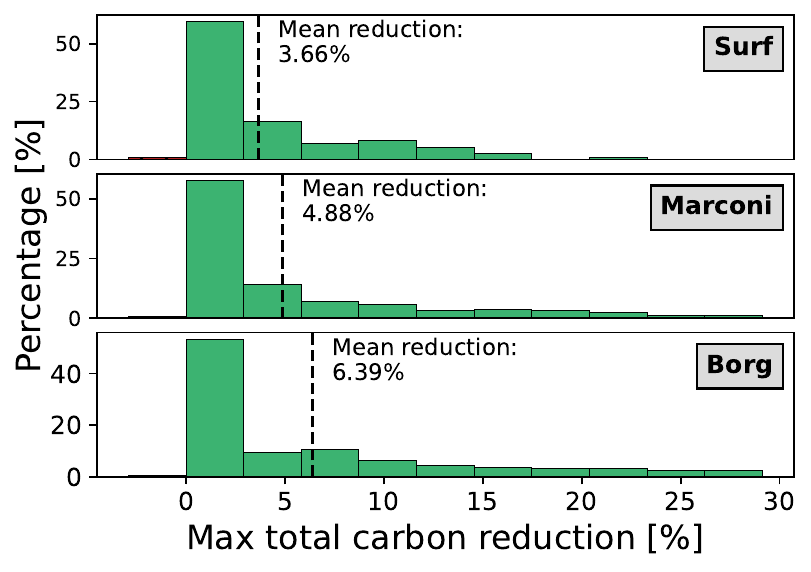}
        \begin{minipage}[h]{0.95\textwidth}
            \centering
            \caption{Total carbon emission reduction using temporal shifting and batteries in 158 carbon regions on three workloads. }\label{fig:TS_BAT_impact}
        \end{minipage}
    \end{minipage}
    \begin{minipage}[t]{0.32\textwidth}
        \centering
        \includegraphics[width=\textwidth]{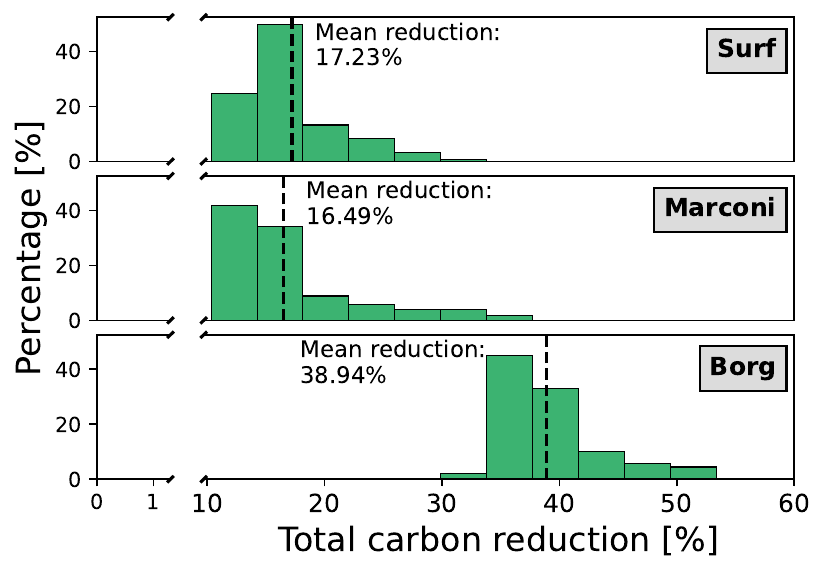}
        \begin{minipage}[h]{0.95\textwidth}
            \centering
            \caption{Total carbon emission reduction using horizontal scaling and battery in 158 carbon regions on three workloads. }\label{fig:HS_BAT_impact}
        \end{minipage}
    \end{minipage}
    \begin{minipage}[t]{0.32\textwidth}
        \centering
        \includegraphics[width=\textwidth]{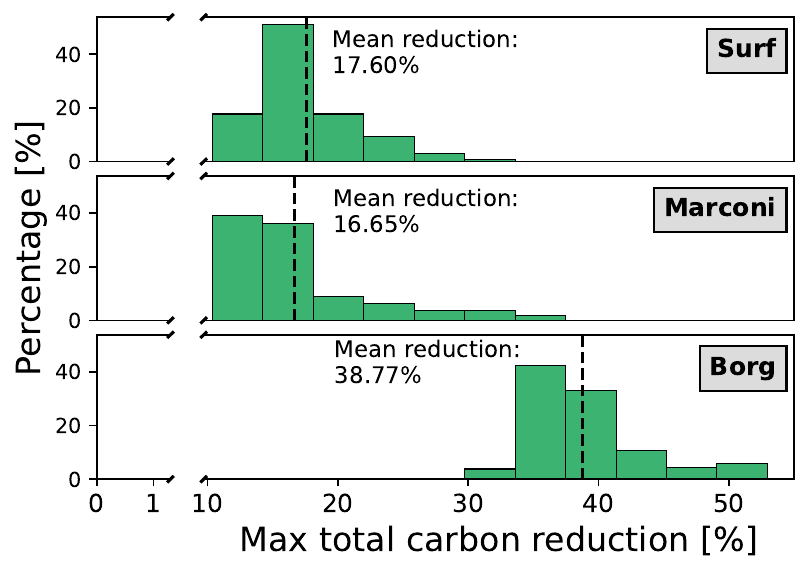}
        \begin{minipage}[h]{0.95\textwidth}
            \centering
            \caption{Total carbon emission reduction using temporal shifting in 158 carbon regions on three workloads.}\label{fig:TS_HS_BAT_impact}
        \end{minipage}
    \end{minipage}
\end{figure*}



\section{Reproducibility Artifact}

\subsection{Paper's Main Contributions}

The main contributions of the paper are the following:

\begin{description}
    \item[$C_1$] We introduce \toolname{}, an open-source datacenter framework that can be used to determine the effectiveness of sustainability techniques on a datacenter's operational and embodied carbon.  
    \item[$C_2$] We execute an extensive analysis of the effectiveness of three sustainability techniques (Horizontal Scaling, Temporal Shifting, and Batteries) in reducing a datacenter's carbon emissions. Besides the carbon reduction, we investigate the potential downsides of using these techniques.  
    \item[$C_3$] We use \toolname{}'s composable design to investigate the effect of using multiple sustainability techniques in conjunction. We show that some techniques are fully capable, increasing their effectiveness, while others do not work well together. We also show that combining techniques introduces new complex trade-offs.
\end{description}

\subsection{Computational Artifacts}
The artifact of STEAM is shared under MIT licence and can be found at:
\begin{itemize}
    \item \url{https://doi.org/10.5281/zenodo.18959863}
    \item \url{https://github.com/atlarge-research/OpenDC-STEAM}
\end{itemize}

\subsection{How to Use}
 \textbf{Before running:} The user should set the base\_folder variable in \texttt{utils/variables.py} to the path to the artifact. This ensures all scripts are using the correct paths.\\
The artifact consists of four sections: 
\begin{enumerate}
    \item \textit{Generating input files: } The artifact provides functions to generate the large number of input files required to run the experiments.
    \item \textit{Running experiments: } The artifact provides a read-to-use version of STEAM to run the generated experiments. 
    \item \textit{Processing output: } The artifact provides functions to process the raw output generated by STEAM into aggregated CSV files containing all required metrics. The artifact contains the raw output for all the experiments run using the SURF workload.
    \item \textit{Visualization: } The artifact provides all functions needed to generate the figures seen in the paper. All the aggregated CSV files required for visualization are contained in the artifact. 
\end{enumerate}

We have provided our results, so each step should be able to be run separately.


\section{Generating and Running simulations}

In this section, we discuss the process of generating input files and using them to run experiments. All artifacts contain \toolname{}, the simulation tool used to run all experiments, and all input files needed to run all experiments of this work. To simplify the experimentation process, \toolname{} is provided as an executable that can be run from the terminal.

\smallskip
\noindent\textit{Expected Results}
\smallskip

Raw output files are created for each experiment executed and placed in the output folder. Each experiment results in five output files:
\begin{itemize}
    \item \texttt{service.parquet} provides general information about the system, such as the number of tasks active. 
    \item \texttt{powerSource.parquet} provides information about the power sources. 
    \item \texttt{battery.parquet} provides information about batteries when used
    \item \texttt{host.parquet} provides information about hosts, such as the CPU utilization. \textit{Note:} Because this file is not needed for all analyses, it will only be exported for specific experiments when needed.
    \item \texttt{task.parquet} provides information about the tasks run. \textit{Note:} Because this file is not needed for all analyses, it will only be exported for specific experiments when needed.
\end{itemize}

Artifact $A_2$ provides scripts that process, aggregate, and visualize the raw output files.

\smallskip 
\noindent\textit{Expected Reproduction Time (in Minutes)} 
\smallskip

Running an experiment takes an average of 1, 4, and 15 minutes for the Surf, Marconi, and Borg workloads used in the paper. For each workload, around 5500 experiments are run. However, \toolname{} is a single-threaded application. Because of this, we could run up to 10 experiments in parallel on a machine. With better hardware, this could be increased even further.

\smallskip 
\noindent\textit{Artifact Setup (incl. Inputs)} 
\smallskip 

\textit{Hardware:} \toolname{} can be used on most Hardware. We advise at least 20GB of available RAM per experiment for larger workloads, such as Borg. For the experiments in this paper, we have used 20 nodes of a scientific cluster, each equipped with a single 24-core 2.8 GHz AMD EPYC-2 (Rome) 7402P CPU and 128 GB of RAM. This allowed us to run up to 10 experiments in parallel on each node.

\textit{Software:} Java 21 is required to use \toolname{}.

\textit{Datasets / Inputs:} All inputs required to run the experiments of this paper are provided. This includes the following folders:
\begin{itemize}
    \item \textit{carbon\_traces} contains carbon traces from 158 regions during 2021-2024 collected from ElectricityMaps\footnote{https://www.electricitymaps.com/}. The traces are converted to parquet files.   
    \item \textit{experiments} contains all the experiment files needed to run the experiments of this work for the three workloads used. A generic function for creating experiment files is provided in \texttt{generate\_experiment.py}. This function is used to create the experiment files needed for this paper in bulk using \\\texttt{generate\_experiments\_borg.py}, \\\texttt{generate\_experiments\_marconi.py}, and \\\texttt{generate\_experiments\_surf.py}.
    \item \textit{failure\_traces} contains the failure traces used in the horizontal scaling experiments.
    \item \textit{topologies} contains all topology files needed to run the experiments in this paper. A generic function for creating topology files is provided in \texttt{generate\_topology.py}. This function is used to generate the topology files required for this paper in bulk using \texttt{generate\_topologies\_borg.py}, \\\texttt{generate\_topologies\_marconi.py}, and \\\texttt{generate\_topologies\_surf.py}.
    \item \textit{workload\_traces} contain the workloads needed to run all experiments. The workloads consist of two files: \texttt{tasks.parquet} defines when tasks arrive and their computational requirements. \texttt{fragments.parquet} defines the computational requirements of each task during execution. All three workloads used in the paper are provided in the artifact.
\end{itemize}

\textit{Installation and Deployment:} \toolname{} is provided as an executable and can thus be run directly from the terminal without any installations needed. 

\smallskip 
\noindent\textit{Artifact Execution} 
\smallskip 

Running experiments is done through the terminal and is very straightforward. From the base folder of the artifact, run the following command: \texttt{./STEAM/bin/STEAMExperimentRunner --experiment-path "path/to/experiment"}. All other input files, such as the topology or the workload, are defined in the experiment file. \textit{Important:} The experiment has to be run from the base folder because we use relative paths to the other files. The output files are placed in the \texttt{output} folder, which will be created if it does not yet exist.\\

\textbf{Parallel Execution:} Because we are running a large number of experiments, and STEAM is a single-threaded application, we recommend executing the experiments in parallel.

\section{Processing and Visualization}
This section discusses all functions used to aggregate, analyze, and visualize the results. 

\smallskip 
\noindent\textit{Expected Results}
\smallskip

The raw output files are processed and combined into aggregated CSV files. The aggregated files are used to generate all experiment figures in the paper.

\smallskip 
\noindent\textit{Expected Reproduction Time (in Minutes)} 
\smallskip  

Processing and combining the raw output files into aggregated CSV files for all workloads takes around 20 minutes. However, this is a single-threaded process that can easily be run in parallel, reducing the runtime significantly. All the figures are created within a minute.

\smallskip 
\noindent\textit{Artifact Setup (incl. Inputs)} 
\smallskip  

\textit{Hardware:} All processing, analysis, and visualization were performed on a \texttt{Lenovo ThinkPad P14s Gen 5} equipped with an \texttt{Intel® Core™ Ultra 7-155H processor} and 64 GB DDR5 RAM. 

\textit{Software:} Processing and analysis are done using Python 3.12.3 with the NumPy, Pandas, and Matplotlib packages. \texttt{PyArrow} is required to be able to read parquet files using \texttt{Pandas}. Finally, we use \texttt{brokenaxes}\footnote{https://pypi.org/project/brokenaxes/} to simplify the process of creating broken axes in Figure 7. 

\textit{Datasets / Inputs:} All required data is provided. The results folder contains the aggregated CSV files generated from the raw output files. The raw output files for the Surf workload are also included as an example (Adding the files for all workloads was impossible due to size constraints). Besides the output files, the \texttt{carbon\_traces} and \texttt{workload\_traces} are required for processing.

\textit{Installation and Deployment:} All scripts can be executed directly from the terminal using Python. No additional installation or deployment is required.

\smallskip 
\noindent\textit{Artifact Execution} 
\smallskip 
\textbf{Processing: }
Before visualizing the output data, it must be organized, processed, and combined.
This is done by running the following files: 
\begin{itemize}
    \item \texttt{process\_output.py} loops over all the raw output files, collects required metrics, and writes them to an aggregated CSV file. 
    \item \texttt{process\_technique\_impact.py} is a tailor-made function that gathers the metrics required for Figure 11. The results of this process are saved as \texttt{surf\_peak\_powers.csv}, \texttt{surf\_task\_delays.csv}, and \texttt{surf\_total\_energies.csv}.  
\end{itemize}

\textbf{Visualization:}
Next, we can visualize the results using the scripts defined in the \texttt{visualization} folder. The \texttt{plotting\_functions} contains all the Python scripts needed to visualize figures 5-12 of the paper. \texttt{plot\_all.sh} can be used to create all figures. \textbf{Note:} The aggregated CSV files required to make all figures are available in the artifact.

\fi

\end{document}